# Bessel Speckles

*Vijayakumar Anand\**

Institute of Physics, University of Tartu, W. Ostwaldi 1, 50411 Tartu, Estonia
E-mail: vijayakumar.anand@ut.ee

Keywords: ((Bessel beams, speckles, incoherent imaging, holography, diffractive optics, imaging))

((Abstract text. Maximum length 200 words. Written in the present tense.))
Speckle patterns are formed by random interferences of mutually coherent beams. While speckles are often considered as an unwanted noise in many areas, they also formed the foundation for the development of numerous speckle-based imaging, holography and sensing technologies. In the recent years, artificial speckle patterns have been generated with spatially incoherent sources using static and dynamic optical modulators for advanced imaging applications. In this report, a fundamental study has been carried out with Bessel distribution as the fundamental building block of the speckle pattern: speckle patterns formed by randomly interfering Bessel beams. Indirect computational imaging framework has been applied to study the imaging characteristics. In general, Bessel beams have a long focal depth, which in this scenario is counteracted by the increase in randomness enabling tunability of the axial resolution between the limits of Bessel beam and a Gaussian beam. Three-dimensional computational imaging has been synthetically demonstrated. The presented study will lead to a new generation of incoherent imaging technologies.

## 1. Introduction

The observation and study of speckle patterns dates back to the 20[th] century. [1,2] A speckle pattern is generated by an interference among mutually coherent optical fields with random amplitude and phase commonly originating from a scattering object. [3] In the beginning, speckle patterns were considered as a problem to be minimized in imaging and many speckle suppression techniques were developed to clean the images of the speckle noises. [4-6] Later, the randomness associated with the speckle patterns was found attractive for many applications such as interferometry, astronomical speckle imaging, optical lever, non-invasive imaging, super resolution imaging, biomedical imaging and 3D sensing. [6-13] In the recent years, artificial speckle patterns have been generated with spatially incoherent light sources using static as well as dynamic optical modulators for holography applications. The main difference between the speckles formed by a coherent and incoherent source originates from the coherence condition. In the case of a coherent source, the light from every object point is scattered and interferes with the scattered light from another point. With an incoherent source, the light from every object point is scattered, whose intensity is added with the scattered intensity of another object point. Consequently, the process of addition in the case of an incoherent source reduces the visibility but allows the linearity condition in intensity which subsequently leads to a simpler imaging process. [14,15]

The first such study of using speckles for incoherent imaging was reported by Dicke and Ables in the year 1968, where a random pinhole array was used as the static optical modulator between the object and sensor planes. [16,17] The indirect imaging method consists of three steps: calibration, imaging and reconstruction. In the calibration step, the point spread function (*PSF*) of the system was recorded. The object intensity was recorded for a test object in the next step. In the final step, the image of the object was reconstructed by processing the *PSF* with the object intensity distribution in the computer. This computational reconstruction process can be as simple as a cross-correlation [18] or a highly complicated algorithm. [19] In the recent years, speckles formed from incoherent light sources have been used for 3D,[20,21] 4D[22] and 5D[23] imaging applications. In many studies, the degree of chaos in the speckle pattern was controlled by controlling the scattering degree. [24,25] However, beyond that variation, there was no parameter that could be varied and in far-field the speckle sizes of weak and strong diffusers become equal resulting in a similar imaging behavior except for slight change in the background noise. Therefore, the imaging characteristics of incoherent imaging systems based on speckles cannot be tailored beyond the limits of direct lens-based imaging system recalling that regular spatial correlation lengths are often diffraction limited. [25] In special cases, it is also possible to obtain correlations lengths beyond the diffraction limits by aperture engineering and the use of advanced reconstruction methods. [26-28]

With the developments in active optical modulators and the rapid developments in the area of imaging technologies, it is necessary to dissect the speckle theory further. In this study, for the first time, Bessel beams were used as the fundamental building block for the generation of speckle patterns. Bessel beams have many interesting characteristics such a self-healing, non-diffracting and most importantly a long focal depth. [29-38] An object point is converted into a random array of interfering Bessel beams with different relative intensities and propagation directions. The sparsity of the interactions was varied and the imaging characteristics were studied. A synthetic computational 3D imaging was carried out using two test objects for a few cases of sparsity conditions. The manuscript consists of five sections. The theoretical analysis is presented in the next section. In the third section, simulation studies are presented. The synthetic computational 3D experiments are presented in the fourth section. In the final section, the conclusion and future perspectives are discussed.

## 2. Methodology

The concept figure for the proposed study is shown in Figure 1. The light from an object point is converted by an optical modulator into a random array of interfering Bessel beams. The first step is to define and understand sparsity, which in this study is quantified by $\sigma$ which is the inverse of number of interfering Bessel beams $N$, when the number of beams is 1, $\sigma = 1$ and when the number of beams approaches a large number, $\sigma \to 0$. [39] When sparsity increases, the randomness decreases and vice versa. In the first step, the optical modulator needs to be designed that can map every object point into a random array of interfering Bessel beams. There are different approaches to achieve this design. One direct approach involves the use of Gerchberg-Saxton algorithm (GSA) with the spot arrangement in the spectrum domain and the phase-only function at the space domain can be calculated. [39,40] It is a well-known that axicons in both refractive and diffractive versions can be used to generate Bessel beam. [41] As this is a theoretical study, the optical modulator was designed by summing diffractive axicon phases with different periods and linear phases. The complex amplitude of the optical modulator is given as

$$\Psi_{OM} = \sum_{p=1}^{N} A_p \exp\left[-j2\pi \left(\frac{R}{\Lambda_{1p}} + \frac{X}{\Lambda_{2p}} + \frac{Y}{\Lambda_{3p}}\right)\right], \quad (1)$$

where, $R = \sqrt{X^2 + Y^2}$ is the radial coordinate, $\Lambda_1$, $\Lambda_2$ and $\Lambda_3$ are the periods of the axicon, $X$ and $Y$ components of the linear gratings respectively, $A$ is the amplitude, $N$ is the number of Bessel beams and $\sigma = 1/N$. The randomness of $A$, $\Lambda_1$, $\Lambda_2$ and $\Lambda_3$ was controlled by an independent uniform random variable $C \cdot U[0, 1]$, where $C$ is a constant used to set limits for different parameters namely amplitude and periods.

The light from a point object with an amplitude of $\sqrt{I_0}$ reaches the optical modulator located at a distance of $z_s$. The complex amplitude entering the optical modulator is given as $C_1\sqrt{I_0}L(\bar{r}_o/z_s)Q(1/z_s)$, where $Q(a) = exp\left[j\frac{\pi a}{\lambda}R^2\right]$ is a quadratic phase function and $L(\bar{o}/u) = exp[j2\pi(o_x x + o_y y)/(\lambda z_s)]$ is a linear phase function. The complex amplitude after the optical modulator is given as $C_2\sqrt{I_0}Q(1/z_s)L(\bar{r_o}/z_s)\Psi_{OM}$. The *PSF* is observed at a distance of $z_h$ from the optical modulator which is given as

$$I_{PSF} = \left|C_2\sqrt{I_0}Q(1/z_s)L(\bar{r_o}/z_s)\Psi_{OM} \otimes Q(1/z_h)\right|^2,$$

(2)

where '$\otimes$' is a 2D convolutional operator. For an axicon, i.e., $\Psi_{OM} = \exp\left[-j2\pi\frac{R}{\Lambda_{1p}}\right]$, there is always an annular region in the axicon that satisfies the imaging condition of Eq. (2) resulting in the central maximum in the sensor plane. The other regions of the axicon that do not satisfy the imaging condition form the rings around the central maximum, larger the radius of the ring, larger is the phase difference between the phase for the imaging condition and the phase of the region. [42] When $z_s$ is varied in Eq. (2), the central maximum is contributed by different annular regions of the axicon until it reaches a point when the phase of any region of axicon can't satisfy the imaging condition. At this point, lies the boundary of the focal depth and the pattern starts to change into a ring pattern becoming larger in diameter with distance henceforth.

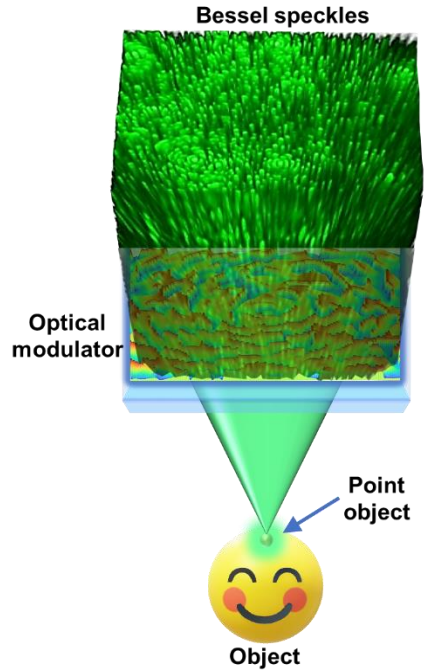

**Figure 1** Optical configuration for generation of Bessel speckles

When the number of Bessel beams increase, then in addition to propagation, there is also interactions between the Bessel beams resulting in self-interference distributions. While every Bessel beam is non-diffracting by nature, with an increase in number of Bessel beams, the interactions increase which changes the intensity distributions in the sensor plane with depth.

These interactions vary randomly as the phase of the different Bessel beams vary with the change in the location of the object point. A single Bessel beam may be applied for imaging application but with more than one Bessel beam, the object information becomes distorted in the sensor plane. Since the study is carried out using spatially incoherent and temporally coherent light, the system is linear in intensity. Therefore, it is possible to study the imaging characteristics in indirect imaging mode. For an object given by a function '$O$' the intensity distribution generated in the sensor plane is given as $I_O = |O \otimes I_{PSF}|$. As long as $I_{PSF}$ is a Delta-like function, the object information gets mapped point-to-point on the sensor. So, in the case of single Bessel beam, the central maximum samples the object information faithfully, while the rings around the central maximum generate imaging noises. In this case and also the cases where the $I_{PSF}$ is not a Delta-like function, indirect imaging method needs to be applied. The image of the object is reconstructed by a non-linear reconstruction (NLR) method given as $I_R = |\mathcal{F}^{-1}\{|\widetilde{I_{PSF}}|^o \exp[j \cdot \arg(\widetilde{I_{PSF}})]|\tilde{I}_O|^r \exp[-j \cdot \arg(\tilde{I}_O)]\}|$, where $o$ and $r$ are varied until the lowest reconstruction noise is obtained, arg(·) refers to the phase and $\tilde{B}$ is the Fourier transform of $B$. So, in the indirect imaging framework, the imaging *PSF* is not a Bessel distribution but a non-linear autocorrelation of a Bessel distribution which is a Delta-like function. [43]

## 3. Simulation studies

A simulation is carried out in MATLAB with a matrix size of 500 pixels along *x* and *y* directions, sampling size of 10 μm and λ = 632 nm. For the simulation study, the sparsity values $\sigma$ = 1, 0.2, 0.1, 0.05, 0.025, 0.0125 and 0.00625 are considered. Like a direct lens-based imaging system, an indirect imaging system also has the same lateral and axial resolution limits given as ~λ/*NA* and ~λ/*NA*$^2$ respectively, where *NA* is the numerical aperture $D/z_s$, where *D* is the diameter of the entrance pupil which is 5 mm and $z_s$ and $z_h$ is 50 cm in this study. To have a reliable comparison, the simulation results are also compared with direct imaging using a Fresnel Zone Plate (FZP) with $z_s$ and $z_h$ in 2*f* configuration (*f* = 25 cm) and an axicon with a minimum period. The amplitude and phase of the optical modulators FZP, diffractive axicon ($\sigma$ = 1) and optical modulators for generating self-interfering Bessel beams for $\sigma$ = 0.2, 0.1, 0.05, 0.025, 0.0125 and 0.00625, and their axial distributions from the optical modulator to the sensor are shown in Figure 2. The axial distribution was calculated at every plane by varying $z_h$ in Eq. (2) from the plane of the optical modulator to the sensor plane and the 2D information was accumulated into a cube data. As seen from Figure 2, with an increase in the sparsity $\sigma$, the density of the Bessel beams increases. The individual Bessel beam can be identified by the green lines while the side lobes and the interference effects are observed in orange color.

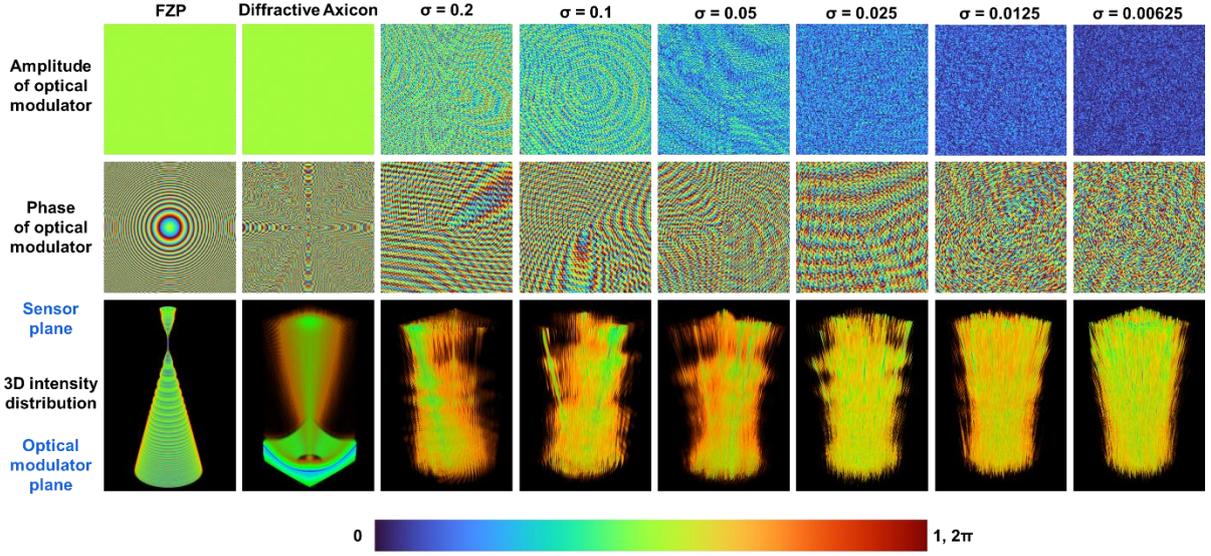

**Figure 2** Amplitude, phase and axial intensity distribution for FZP, diffractive axicon ($\sigma = 1$), self-interfering Bessel beams $\sigma = 0.2, 0.1, 0.05, 0.025, 0.0125$ and $0.00625$.

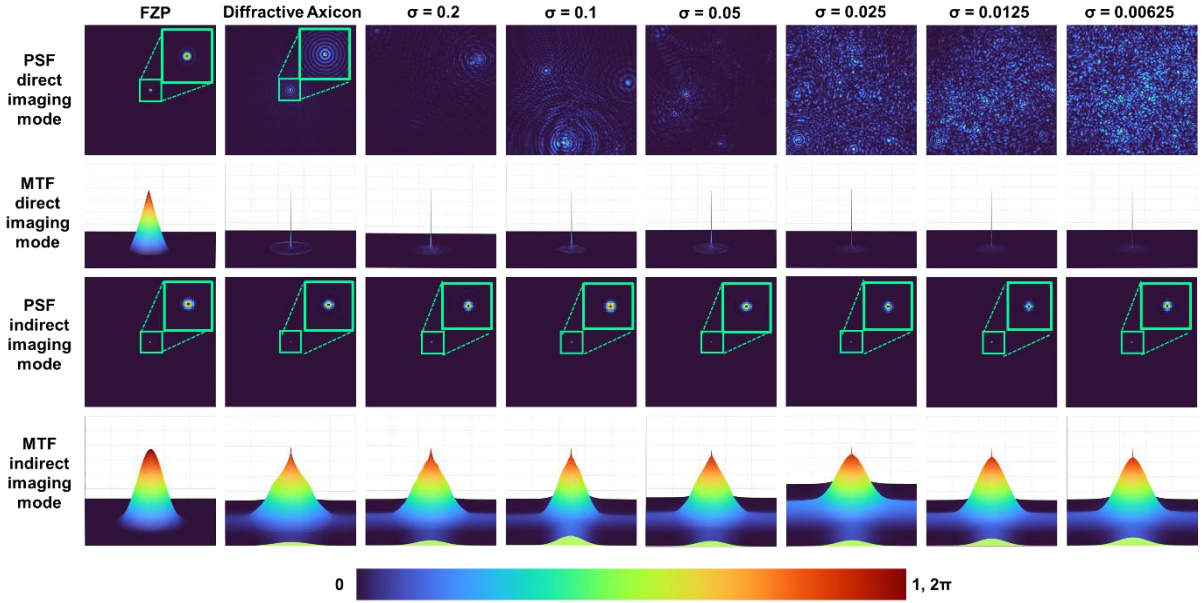

**Figure 3** PSF and MTF obtained for direct and indirect imaging modes for FZP, diffractive axicon ($\sigma = 1$), self-interfering Bessel beams $\sigma = 0.2, 0.1, 0.05, 0.025, 0.0125$ and $0.00625$. The MTFs are normalized to 1.

The images of the *PSF*s for FZP, axicon and optical modulators with $\sigma = 0.2, 0.1, 0.05, 0.025, 0.0125$ and $0.00625$ at the sensor plane located at 50 cm from the optical modulator and their corresponding modulation transfer functions (MTF) given as $\mathbf{MTF} = a|\mathcal{F}(\mathbf{I_{PSF}})|$, indirect

*PSF*s obtained using NLR and their MTFs are shown in Figure 3. The optimal values of *o* and *r* in this case was 0 and 0.6 respectively. It is seen that in indirect imaging mode, the number of Bessel beams have almost no impact on the lateral resolution of imaging. This is an interesting result as the increase in randomness significantly changes the intensity distribution in direct imaging mode, while it is not varying in indirect imaging mode. The next important characteristic of an imaging system is its' axial resolution. The axial resolution is the measure of how rapidly the intensity along the optical axis decreases when the object location is shifted away from the imaging condition. In Indirect imaging mode, this can be measured by a cross-correlation between the intensity distribution corresponding to a particular plane in the object space and the one obtained corresponding to another plane. The axial intensity variation curve can be obtained as $I_a = \left|\mathcal{F}^{-1}\left\{\left|\widetilde{I_{PSF}(\Delta z = 0)}\right|^o \exp[j \cdot \arg(\widetilde{I_{PSF}(\Delta z = 0)})]\left|\widetilde{I_{PSF}}(\Delta z)\right|^r \exp[-j \cdot \arg(\widetilde{I_{PSF}(\Delta z)})]\right\}\right|$, where $\Delta z$ is the shift in the location of the object in the object plane. The point object's location was shifted from 25 cm to 75 cm and the axial curve was calculated for $\sigma = 1$ to 0.00625 as shown in Figure 4. It can be seen that with an increase in the number of Bessel beams, the focal depth decreases in the indirect imaging mode. This is an anomaly or non-linearity occurring only in the indirect imaging mode. In a direct imaging mode, the axial resolution and lateral resolution are intertwined and with a change in one affects the other. Let us consider a classical experiment of decreasing the aperture diameter. In this case, when the lateral resolution decreases, the axial resolution decreases and vice versa. However, in indirect imaging mode, the lateral resolution is independent of the number of interfering Bessel beams while the axial resolution increases with an increase in the number of interfering Bessel beams. This is a unique behavior and will benefit imaging applications.

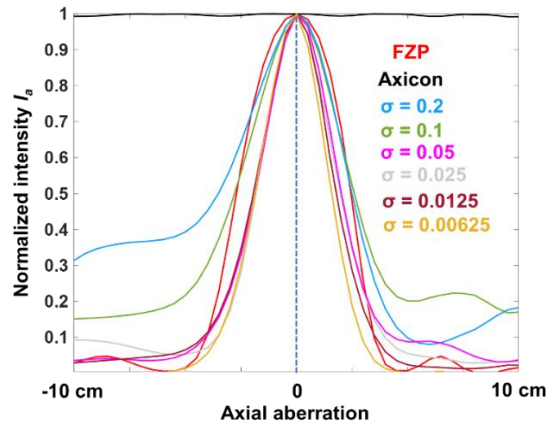

**Figure 4** Plot of $I_a$ with respect to $\Delta z$ for FZP, axicon and optical modulator with $\sigma$ = 0.2, 0.1, 0.05, 0.025, 0.0125 and 0.00625.

In the past, such hybridization was achieved by complicated holography experiments [44]. It must be noted that in all the reported imaging systems based on speckles such as interferenceless coded aperture correlation holography [20], DiffuserCAM [19] and scatterplate microscope [21] such a non-linearity was not observed. In this case, the fundamental building block of the speckle pattern has been modified to introduce such a non-linearity where the lateral resolution is constant but the axial resolution varies for different cases.

## 4. Three-dimensional Imaging

A synthetic three-dimensional (3D) imaging is carried out next using two objects namely "CIPHR" and "Annalen der Physik" mounted at two different planes separated by a distance of 25 cm. The *PSF*s corresponding two planes and the imaging results obtained by refocusing at the two planes for direct imaging with FZP and a diffractive axicon ($\sigma = 1$) are shown in Figure 5. As seen in the figures, the PSFs of a diffractive axicon remain unchangeable even with an axial error of 25 cm, while the PSFs of an FZP changes significantly. Consequently, the object information of one plane in the case of FZP is highly distorted, while in the case of diffractive axicon due to a high focal depth the information from both planes are perceived. However, the Bessel intensity distribution which is the reason for the high focal depth also distorts the information due to the ring pattern around the central maxima. Comparing the object information in the imaged planes, the object information obtained for FZP is sharper than that of diffractive axicon.

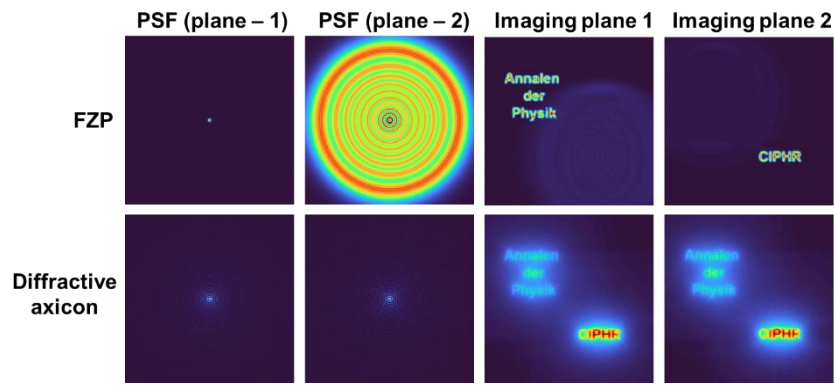

**Figure 5** Images of PSFs and the imaging results of the two planes with two test objects for FZP and diffractive axicon.

The PSFs, object information and the reconstruction results at the two planes corresponding to the two PSFs for $\sigma = 0.2, 0.1, 0.05, 0.025, 0.0125$ and $0.00625$ are shown in Figure 6. As per the linearity condition in intensity, the object intensity distributions for object 1 and object 2 were simulated by a convolution operation with the corresponding PSFs. The resulting intensity distributions are summed to form the hologram. The object hologram can be mathematically constructed as $H = \sum_{q=1}^{2}|I_{PSF}(z_q) \otimes O_q|$. It can be seen that with a decrease in the sparsity, the axial resolution improves. All the reconstructions were carried out for $o = 0$ and $r = 0.6$. It is seen for σ = 0.2, the other plane information is quite visible in both reconstructions. As σ gradually decreases, the other plane information becomes less visible which indicates the increase in the axial resolution.

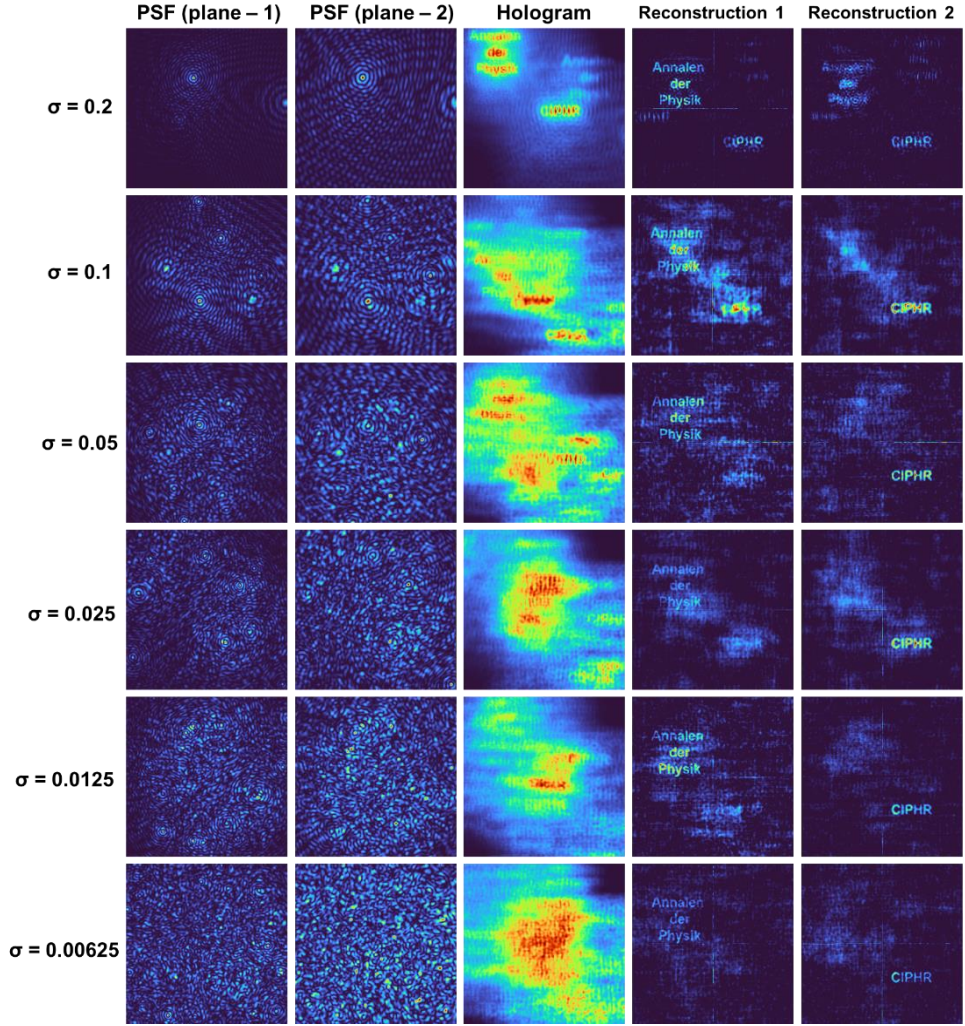

**Figure 6** Images of PSFs at two planes separated by a distance of 25 cm, hologram and reconstruction results corresponding to the two planes $\sigma$ = 0.2, 0.1, 0.05, 0.025, 0.0125 and 0.00625.

## 5. Conclusion

Speckle patterns are formed by the random interferences of mutually coherent waves originating from diffusive surfaces. Speckles, once considered a noise to be suppressed have led to the development of numerous imaging, sensing, interferometry and holography technologies. In the recent years, there has been numerous studies reported on the application of artificial speckles formed by spatially incoherent light for 3D, 4D and 5D imaging and holography applications. [45-48] In this study, for the first time, the fundamental building block of speckle pattern has been selected as a Bessel beam and the imaging characteristics have been studied in an indirect imaging framework. In a classical imaging system, the lateral and axial resolutions are dependent on one another, where it is almost impossible to vary one without affecting the other. In this study, where Bessel beams were randomly interfered in the indirect imaging framework, a non-linear behavior was observed. When the number of self-interfering Bessel beams was increased, the lateral resolution remained a constant, while the axial resolution increased. This is a useful behavior for imaging applications.

A synthetic 3D imaging has been carried out using two test objects following the principles of intensity linearity in imaging and a similar behavior was observed. There is a generation of background noise which can be suppressed using statistical averaging with intensity distributions formed from mutually exclusive aperture configurations. This study is a significant step towards generalizing the concepts of indirect imaging using spatially incoherent light. In this case, two parameters namely the non-diffraction parameter of Bessel beam and the accumulation of random phases associated with the Bessel beams compete against one another creating a possibility to tune the axial resolution of the system without altering the lateral resolution. While in this study, only axial coordinate has been considered, a similar effect is also expected along the spectral coordinate. Bessel beam's intensity distribution remains constant with a change in wavelength but with the ability counteract the spectral behavior with randomness, it is possible to control the spectral resolution of the system. In the same framework, the study of polychromatic Bessel beams will be another interesting direction. In summary, I believe that this study will lead to a new generation of incoherent imaging and holography technologies, where the ingredients of an ensemble of self-interfering beams can be engineered to achieve desired imaging characteristics.

**Acknowledgements**
The author acknowledges the European Union's Horizon 2020 research and innovation programme grant agreement No. 857627 (CIPHR). The author thanks Ms. Tiia Lillemaa for the administrative support.

**Data availability**

The data that support the findings of this study are available from the corresponding author upon reasonable request.

**Conflict of interests**

The author does not have any conflict of interest.

**ToC**

Artificial speckle patterns have been constructed with Bessel beam as the fundamental building block and studied in a spatially incoherent indirect imaging framework. The non-diffracting behavior of Bessel beams competes with the randomness allowing to tune the axial resolution of the system without changing the lateral resolution of the system.

V. Anand*

**Bessel Speckles**

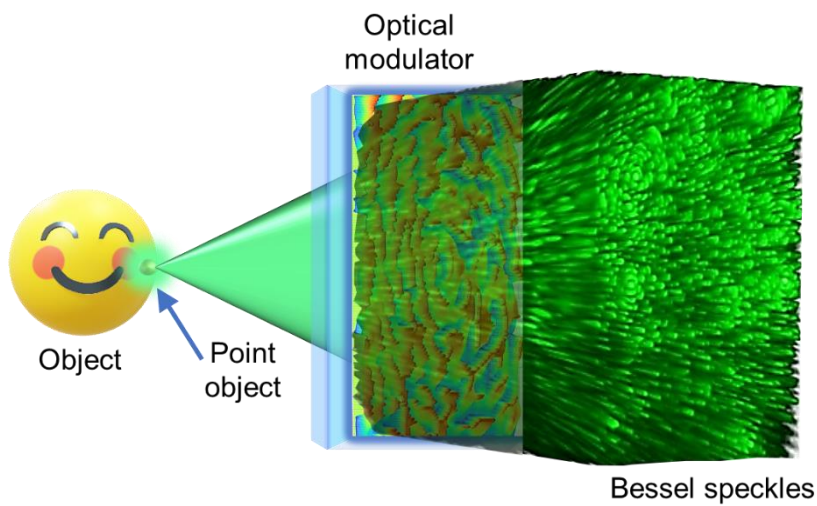